# Magnet-free nonreciprocal metasurface for on-demand bi-directional phase modulation

Weihao Yang[1,2], Jun Qin[1,2*], Jiawei Long[2], Wei Yan[1,2], Yucong Yang[1,2], Chaoyang Li[1,2,5], En Li[2], Juejun Hu[3], Longjiang Deng[1,2*], Qingyang Du[3,4*] and Lei Bi[1,2*]

[1]National Engineering Research Center of Electromagnetic Radiation Control Materials, University of Electronic Science and Technology of China, Chengdu 610054, People's Republic of China

[2]School of Electronic Science and Engineering, University of Electronic Science and Technology of China, Chengdu 610054, People's Republic of China

[3]Department of Materials Science and Engineering, Massachusetts Institute of Technology, Cambridge, Massachusetts, 02139, USA

[4]Department of Electrical Engineering, Xiamen University, Xiamen, Fujian, China 361005

[5]Hainan University, Haikou, Hainan Province 570228, China

qinjun@uestc.edu.cn, denglj@uestc.edu.cn, qydu@xmu.edu.cn, bilei@uestc.edu.cn

**Unconstrained by Lorentz reciprocity, nonreciprocal metasurfaces are uniquely capable of encoding distinctive optical functions on forward- and backward-propagating waves. The nonreciprocal metasurfaces reported to date require external electric or magnetic field biasing or rely on nonlinear effects, both of which are challenging to practically implement. Here, we propose and experimentally realize a magnet-free, linear, and passive nonreciprocal metasurface based on self-biased magnetic meta-atoms. Record transmittance up to 77% and operation angle reaching $\pm 64°$ are experimentally demonstrated. Moreover, on-demand bidirectional phase modulation in a "LEGO-like" manner is theoretically proposed and experimentally demonstrated, enabling a cohort of nonreciprocal functionalities such as microwave isolation, nonreciprocal beam steering, nonreciprocal focusing, and nonreciprocal holography. The design can also be extended to MHz and optical frequencies, taking advantage of the wide variety of self-biased gyrotropic materials available. We foresee that the nonreciprocal metasurfaces demonstrated in this work will have a significant practical impact for applications ranging from nonreciprocal antennas and radomes to full-duplex wireless communication and radar systems.**

Metasurfaces have attracted significant research interest in recent years because they allow versatile control of electromagnetic wave propagation with a flat, low-profile form factor. Their planar structure facilitates scalable fabrication of metasurfaces based on printed circuit boards (PCB) or silicon photonic technologies. Considerable progress has been made in realizing functionalities such as polarization conversion[1-4], perfect absorbtion[5, 6] and wavefront control[7-10]. However, most metasurfaces demonstrated to date obey the Lorentz reciprocity. Nonreciprocal metasurfaces (NRMs) have only recently emerged as a new type of metasurfaces only recently[11-15]. Unlike conventional metasurfaces whose characteristics are symmetric with respect to the wave propagation direction, NRMs are uniquely capable of performing direction-dependent functions. For instance, NRM-integrated radomes can act as nonreciprocal antennas, which can not only curtail antenna echoes[16] but also enable arbitrary emission and receiving characteristics of electromagnetic waves. NRMs also facilitate full-duplex wireless communications by doubling the bandwidth of wireless communications[17].

Several mechanisms can be used to achieve nonreciprocity[18]. A transistor-loaded circuit allows unidirectional gain of electromagnetic waves owing to its rectifying effect, thereby breaking Lorentz reciprocity[19, 20]. These devices can exhibit a broad operational bandwidth and forward transmission gain. However, they suffer from low operation frequencies (limited by transistor speed), low power-handling capability, high biasing power, and poor signal-to-noise ratio[17, 21, 22]. Nonlinearity in metasurfaces with asymmetric structures is another means of generating nonreciprocity, although they are sensitive to incident power and are limited by dynamic reciprocity to pulsed-wave applications only[23, 24]. Reciprocity can also be lifted by spatiotemporal modulation, although this approach incurs a large constant power consumption, a trade-off between modulation frequency and operation bandwidth, and undesired high-harmonic generation[12, 17, 25]. Finally, magnetic materials are naturally nonreciprocal given their asymmetric permittivity or permeability tensors. Even though these materials are widely used in commercial isolators or circulators, they are surprisingly less explored for NRM applications, likely because bulky biasing magnets preclude the local control of individual meta-atoms. To summarize, bias-free, linear, and passive NRMs are lacking for applications in free-space microwave or photonic systems.

Here, we theoretically propose and experimentally demonstrate a new class of NRM consisting of self-biased magnetic meta-atoms. Using all-dielectric Mie resonators made of M-type La-doped $BaFe_{12}O_{19}$ hexaferrite (La:BaM), the local magnetization direction can be locked by the strong magnetocrystalline anisotropy of this material, which eliminates the requirement of magnetic field biasing. Moreover, each meta-atom can be individually magnetized along arbitrary directions, providing unprecedented design flexibility to attain bidirectional amplitude and phase profiles. Following a nonreciprocal digital coding metasurface (NDCM) design methodology, we designed and fabricated NRMs in a "LEGO-like" manner with a diverse set of functions, such as unidirectional transmission, nonreciprocal beam deflection, nonreciprocal beam focusing, and nonreciprocal holography in the $K_u$ band. This NRM platform represents a new class of free-space nonreciprocal devices ideally poised for applications such as electromagnetic wave isolation, circulation, nonreciprocal antennas and radomes, and full-duplex transmission.

**Device structure and operation principles**

The meta-atom of the NRM is shown in Fig. 1a. We chose La:BaM as the magnetic material because of its large off-diagonal component of the permeability tensor (see details in Supplementary Fig. S1) and the high remanent magnetization ($M_r$) resulting from its strong magnetocrystalline anisotropy field ($H_a$ = 1.46 T). The meta-atom consists of a subwavelength La:BaM cuboid pillar on top of a Teflon substrate, with $M_r$ along the surface normal ($z$ or $-z$) direction, as depicted in Fig. 1a. The magnetic hysteresis, permittivity, and permeability tensor of La:BaM in the remanence state are measured and calculated in Supplementary Note 1. In this configuration, a circular polarized electromagnetic wave "sees" different refractive indices and extinction coefficients when the wavevector is parallel or antiparallel to the magnetization direction, which is attributed to magnetic circular birefringence (MCB) and magnetic circular dichroism (MCD) effects (for low loss frequencies MCB dominates). Therefore, the magnetic Mie resonators exhibit different resonance frequencies for the forward and backward propagating waves, leading to an asymmetric transmission amplitude (Fig. 1b) and phase (Fig. 1c). Fig. 1d shows a schematic diagram of the nonreciprocal phase gradient metasurface design, with the orange color indicating $M_r$ along the $+z$ direction and the purple color indicating $M_r$ along the $-z$ direction. The arrangement of these cells can generate arbitrary direction-dependent phase and amplitude profiles by leveraging MCB/MCD effects.

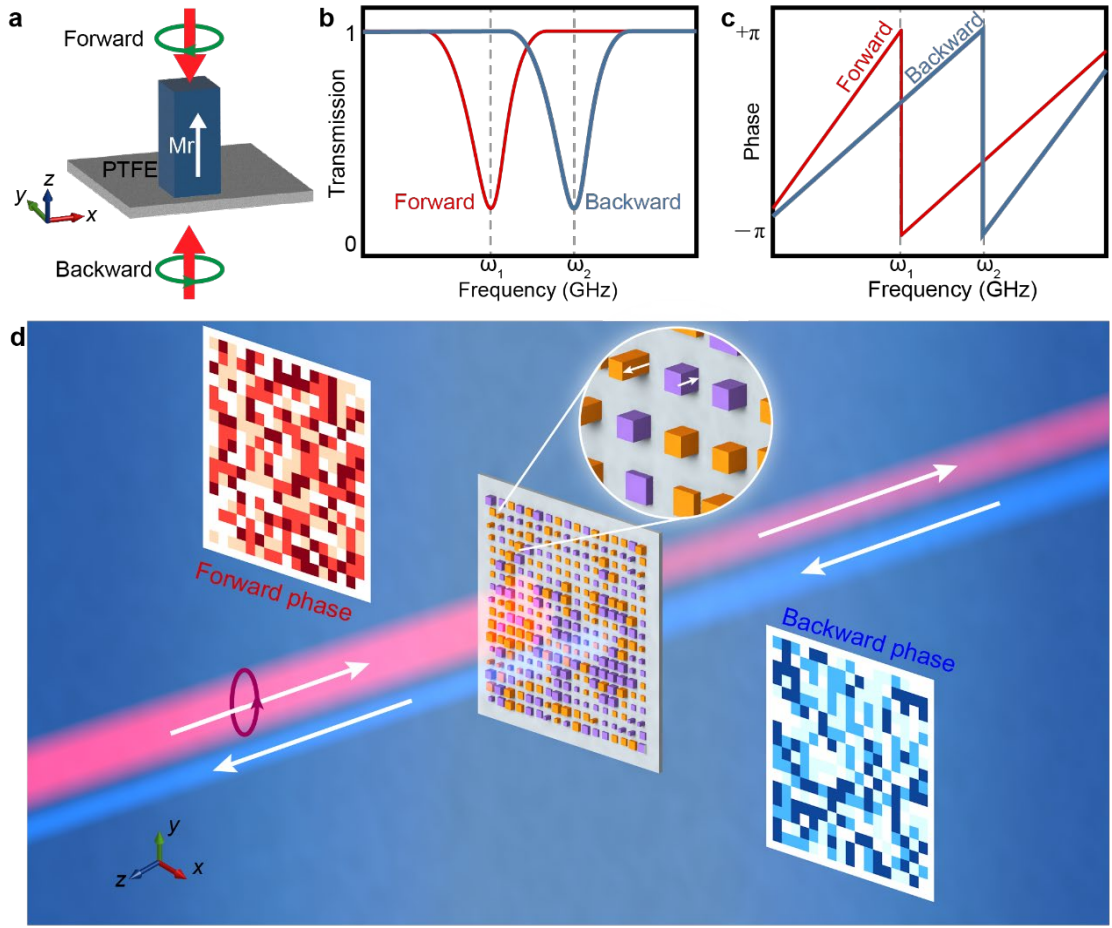

**Fig. 1| Magnet-free nonreciprocal metasurfaces. a**, Operation mechanism of a self-biased magnetic meta-atom. **b**, Nonreciprocal transmission near Mie resonance modes, leading to nonreciprocity in transmission amplitude. **c**, Nonreciprocal transmission near Mie resonance modes, leading to nonreciprocity in transmission phase. **d**, Schematic diagram of the phase gradient nonreciprocal metasurface design principle.

## One-way transmission NRMs

The concept of a one-way transmission NRM is illustrated in Fig. 2a. This device was realized by detuning the Mie resonance wavelength owing to the MCB effect. The metasurface consists of periodically arranged cuboid pillars of La:BaM, which are magnetized under 2.4 T magnetic field and left at remanence before attaching to a 20 cm× 20 cm× 0.2 cm Teflon substrate. The device was designed to operate in the $K_u$ band (~15 GHz, $\lambda = 2$ cm). The width and height of each cuboid were 3 mm and 6 mm, respectively. The period of the meta-atoms was 1 cm. Fig. 2b shows the fabricated sample. We first simulated the transmission spectra of the metasurface under right circular polarized (RCP) incidence for forward and backward transmissions, as shown in Fig. 2c. The forward transmission spectrum shows two resonance peaks at 13.2 GHz and 14.6 GHz, respectively. The two peaks shifted to higher frequencies at 13.6 GHz and 15.4 GHz respectively for backward transmission. Multipole decomposition indicates that the two resonances correspond to a magnetic dipole resonance (MD) mode at 13.2 GHz and a hybrid electric-dipole magnetic-quadrupole (ED-MQ) mode at 14.6 GHz, respectively (see details in Supplementary Fig. S2). Nonreciprocal one-way transmission with an isolation ratio of 10.9 dB and insertion loss of 0.6 dB at 15.4 GHz, and an isolation ratio of 10.4 dB and insertion loss of 5 dB at 13.6 GHz were theoretically predicted. Fig. 2d shows the experimental transmission spectra of the metasurface (see the measurement setup details in Supplementary Fig. S3). An isolation ratio of 7.0 dB and insertion loss of 8.4 dB at 13.7 GHz, and an isolation ratio of 5.6 dB and insertion loss of 1.1 dB at 15.7 GHz were experimentally observed, consistent with the simulation results. The minor difference between the experiment and simulation was likely attributable to the imperfect circular polarization of the source antenna with an axis ratio of 2.5 dB. We further simulated (Fig. 2e) and measured (Fig. 2f) the transmission contrasts at incident angles ranging from 0° to 64°. The results showed that the resonance frequencies (especially those of the ED-MQ mode) were almost independent of the incident angle. The maximum angle was only limited by the finite size of the NRM and our experimental set-up. The results would also be identical for 0° to -64° incidence for symmetry considerations. This wide-angle tolerance suggests that such NRM is compatible with integration on curved surfaces, for instance, on antenna radomes.

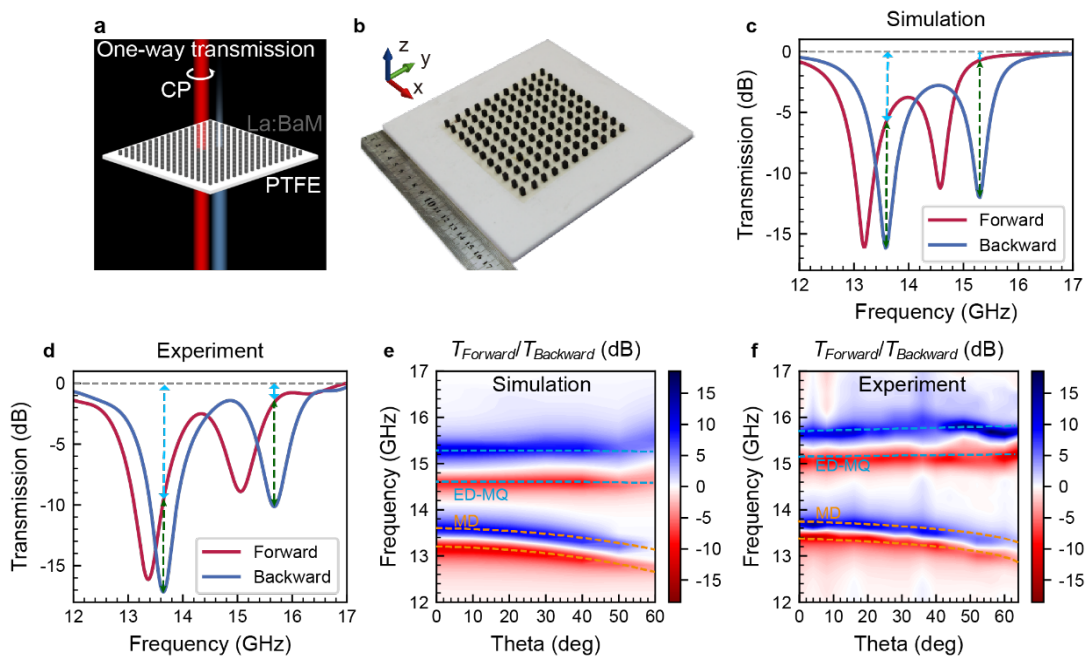

**Fig. 2| A one-way transmission NRM. a**, Schematic diagram of the one-way transmission

NRM. **b**, Image of the fabricated sample. **c**, Simulated and **d**, experimental transmission spectra for forward and backward RCP incidence. **e**, Simulated and **f**, experimental transmission difference between forward and backward RCP incidence as a function of incident angles and frequencies.

**Phase-gradient NRMs**

Phase-gradient NRMs, whose phase profiles differ for forward and backward incidence, allow full on-demand control of bidirectional electromagnetic wave propagation. The design of such metasurfaces is based on the concept of NDCM: each element has both forward and backward propagation phases, which can be written as ($\phi_f(\omega)$, $\phi_b(\omega)$), where $\phi_f(\omega)$ and $\phi_b(\omega)$ are the forward and backward transmission phases, respectively, at frequency ω. To construct a 2-bit phase gradient NRM for arbitrary forward and backward propagation phase profiles, 16 elements were required[26] (Supplementary Fig. S4a). However, because the magnetic meta-atoms can be placed with their magnetization up or down, the phase of the forward and backward incidence can be switched by flipping the direction of $M_r$. This reduces the library elements to only 10 for arbitrary and nonreciprocal phase profiles, as detailed in Supplementary Notes 4 and 5. The optimal meta-atom design must minimize the phase error while maximizing the transmission efficiency for both forward and backward waves. To balance the trade-off between phase error and transmission efficiency, we used a single figure-of-merit to evaluate the meta-atom designs[27] (see details in Supplementary Note 4). The 10 elements were selected in a parameter space consisting of 9,192 parameter combinations (Supplementary Note 4). The transmission spectra of each of the ten fabricated elements were characterized. The transmittance and phase as a function of frequency were characterized by measuring the S21 and S12 parameters, which agreed well with the simulation results presented in Supplementary Note 5. The optimal 10 elements yielded an average transmittance of 42.4% and 38.5%, and an average phase error of ±11.9° and ±12.0° for forward and backward incidence, respectively.

**Nonreciprocal beam deflector**

After finding the 10 nonreciprocal elements, it is possible to construct on-demand bi-directional phase profiles for the NRM in a "LEGO-like" manner. We demonstrated this by first assembling a nonreciprocal beam deflector using meta-atoms (Fig. 3b). Fig. 3a shows the concept of the device. At 14.6 GHz, the device allows forward normal transmittance for the RCP with plane wave incidence, whereas it diffracts the incident wave away from the surface normal for the backward wave. The entire device area measured 17 cm × 10 cm. Four elements with (forward, backward) phase delays of (0, 0), (0, π/2), (0, π), and (0, 3π/2) constituted one period of 10 mm. Therefore, the metasurface exhibited an almost flat phase distribution for forward transmission and a sawtooth phase profile for backward transmission, as shown in Figs. 3c and 3d. The experimentally assessed phase map (stars) agreed well with the target phase profile (dashed lines). Figs. 3e and 3f show the simulated electric field profile at 14.6 GHz, confirming the proposed nonreciprocal diffraction functionality. In our simulation, the diffraction angle reached 29°, with a diffraction efficiency of 17.7% for backward incidence. For forward incidence, the transmission efficiency reached 33.5%. The relatively low transmission efficiency was attributed to the large impedance mismatch at the air/metasurface interface as well as the energy loss into different diffraction orders (<4%). These imperfections induce ~24% and ~34% reflected energy for forward and backward incidence, respectively (see Supplementary Fig. S7). Experimentally, we demonstrate a 31.3% transmission efficiency for the $0^{th}$ order diffraction for forward incidence, and 13.8% transmission efficiency for the $1^{st}$ order diffraction at 30° ± 0.5° for backward incidence. As shown in Figs. 3g and 3h, the radiation patterns exhibited excellent agreement between the experiment and simulation (see experimental details in Supplementary Fig. S8). The slightly lower measured transmission efficiency compared with the simulation may be attributed to the lower transmittance of the experimental meta-atoms and the spherical wavefront of the incident wave (see details in Supplementary Information Fig. S9).

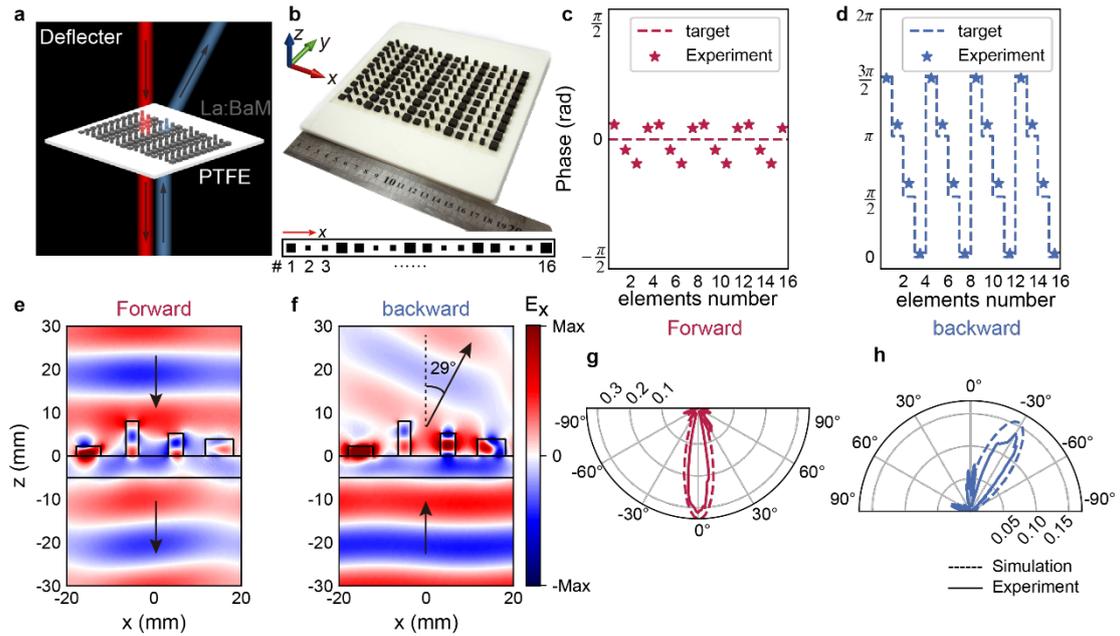

**Fig. 3| Nonreciprocal beam deflector. a**, Schematic of the nonreciprocal beam deflector. **b**,

Image of the fabricated nonreciprocal beam deflector. **c**, Target (dotted line) and measured (star) phase profiles for forward incidence. **d**, Target (dotted line) and measured (star) phase profiles for backward incidence. **e, f**, Simulated electric field profiles of the supercell for **e**, forward incidence and **f**, backward incidence. **g, h**, Measured and simulated far-field radiation patterns for **g**, forward incidence and **h**, for backward incidence.

**Nonreciprocal metalens**

The nonreciprocal metalens exhibit different focal lengths for forward and backward transmission (Fig. 4a). Fig. 4b shows a photograph of the fabricated nonreciprocal metalens sample. The sample area was 17 cm × 17 cm. The meta-atoms were arrayed with a 1 cm period. The hyperbolic phase profiles of the metalens are given by:[28]

$$\varphi_{f/b}(r) = -\frac{2\pi}{\lambda}(\sqrt{r^2 + f_{1/2}^2} - f_{1/2}) \quad , \qquad (1)$$

where $r$ is the radial coordinate, $\lambda$ is the designed wavelength, $f_{1/2}$ is the designed focal length for forward and backward incidence, and $\varphi_{f/b}(r)$ is the phase profile at position $r$ for forward and backward incidence. In our prototype, focal lengths of $f_1 = 6$ cm and $f_2 = 13$ cm were implemented for forward and backward transmission, respectively. The ten elements with proper forward and backward transmission phases were mapped onto the targeted phase profiles of the metalens, as shown in Figs. 4c and 4d. Fig. 4e displays the electric field intensity profiles in the *y-z* plane for forward RCP incidence at 15 GHz. The simulated focal length was 5.4 cm with a focusing efficiency of 46.9%. The discrepancy between the simulated (6 cm) and measured (5.4 cm) focal lengths stemmed from coupling between adjacent meta-atoms (see details in Supplementary Fig. S10). For backward incidence, the focal length and focusing efficiency were 11 cm and 42.3%, respectively. The focusing efficiency was defined as the ratio of the power on the focal plane within a radius of 3×FWHM of the focal spot to the total incident power. The electric field intensity profiles (Figs. 4g and 4h) were measured using a near-field scanning setup via antenna scanning in the *x-y* plane at different *z*-positions (see details in Supplementary Fig. S11). The shadowed areas in both figures could not be probed because of the finite size of the scanning antenna. The intensity profiles showed clear focal spots with focal lengths of 55 ± 1.5 mm and 105 ± 1.5 mm for forward and backward transmission, respectively. The ripple patterns were caused by interference between the incident and reflected waves[29]. The measured focusing efficiencies were 30.5% and 30% for forward and backward incidence, respectively. The lower focusing efficiency in the experiment compared to the simulation was attributed to the spherical instead of the plane wavefront of the incident wave (~7% and

~3% focusing efficiency attenuation for forward and backward incidence, respectively; see details in Supplementary Note 8), as well as the lower transmission of experimentally fabricated meta-atoms compared to simulation (~9% lower transmittance on average, see details in Supplementary Table 1). Fig. 4i shows the line scans of the normalized intensity distributions at the focal plane. The insets of Fig. 4i show the simulated and measured focal-spot images. The dashed curve shows the focal-spot profile of an ideal aberration-free lens with the same aperture size and focal length. The Strehl ratio of the experiment reached 0.81, indicating diffraction-limited focusing of our metalens. The focal spot profiles of backward incidence are presented in Fig. 4j. In this case, the Strehl ratio reached 0.91. The evolution of the focal spot profiles in the $x$-$y$ plane along the $z$-axis was also measured, as shown in Supplementary Fig. S12.

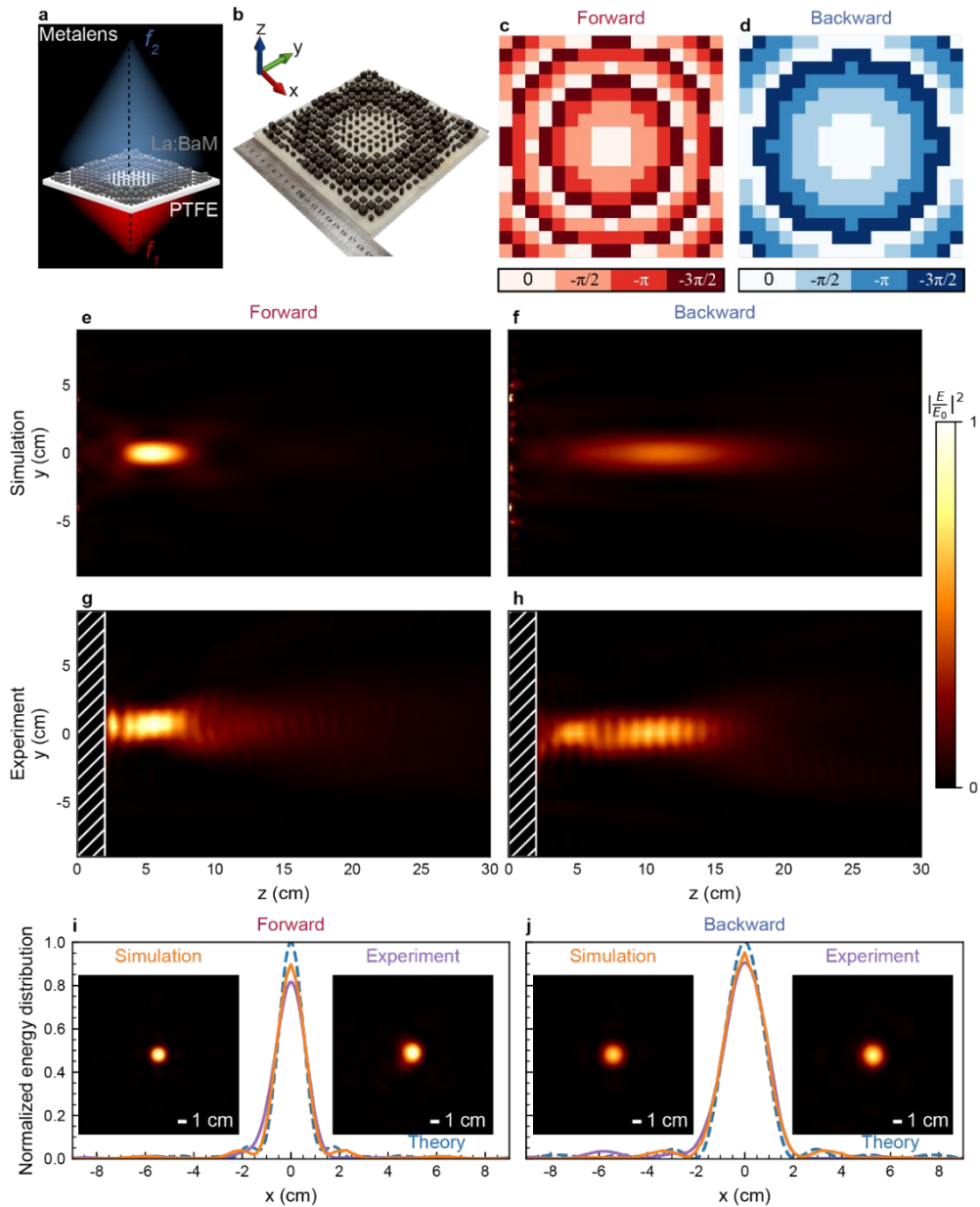

**Fig. 4| Nonreciprocal metalens. a**, Schematic of the nonreciprocal metalens. **b**, Image of the fabricated nonreciprocal metalens. Phase profiles of the metalens for **c**, forward incidence and **d**, backward incidence. Simulated focal spot profiles in the *y-z* plane for **e**, forward incidence and **f**, backward incidence at 15 GHz. Measured focal spot profiles in the *y-z* plane for **g**, forward incidence and **h**, backward incidence. Normalized focal spot distributions along *x*-direction in the focal plane for **i**, forward and **j**, backward incidence.

**Nonreciprocal holography**

Holography is a promising technology for three-dimensional displays[30], beam shaping[31] and artificial intelligence[32]. Here, we utilize the NRM to achieve nonreciprocal holography, which displays different holograms for forward and backward propagations. To design phase-only computer-generated holograms, we selected two image planes 10 cm from the surface of each side of the sample. Hand-written Greek letters "ε" and "μ" were taken as the holographic patterns, as illustrated in Fig. 5a. The Gerchberg–Saxton algorithm was used to retrieve the phase profiles of the hologram (Supplementary Fig. S13). Figs. 5c and 5d depict the discrete phase profiles of the "ε" and "μ" patterns for forward and backward incidences at 15 GHz, respectively. Figs. 5e and 5f show holographic images simulated with the phase profiles in Fig. 5c and 5b, showing clear "ε" and "μ" letters in the respective image planes. Using the same measurement setup used for metalens characterization, intensity profiles of "ε" and "μ" letters (Figs. 5g and 5h) were mapped, confirming good agreement with the design. The overall efficiencies[33, 34] (the fraction of the incident energy that contributes to the transmitted holographic image) for the two incident directions were 27.0% and 26.8%, respectively. Compared to the simulated efficiencies of 44.0% and 42.4%, the difference was caused by the spherical wave front (~5.3% and ~4% deviation from the plane wave), lower transmittance of the meta-atoms in the experiment (~9% lower than the simulation in average), and phase errors of the meta-atoms (~3% deviation from the simulation). In addition, because there were only 17 × 17 meta-atoms in this demo, the energy outside the holographic images could not be completely suppressed, limiting the maximum efficiency to 75% (see details in Supplementary Note 12). These considerations can be improved in future studies to achieve nonreciprocal holography with an improved resolution and efficiency.

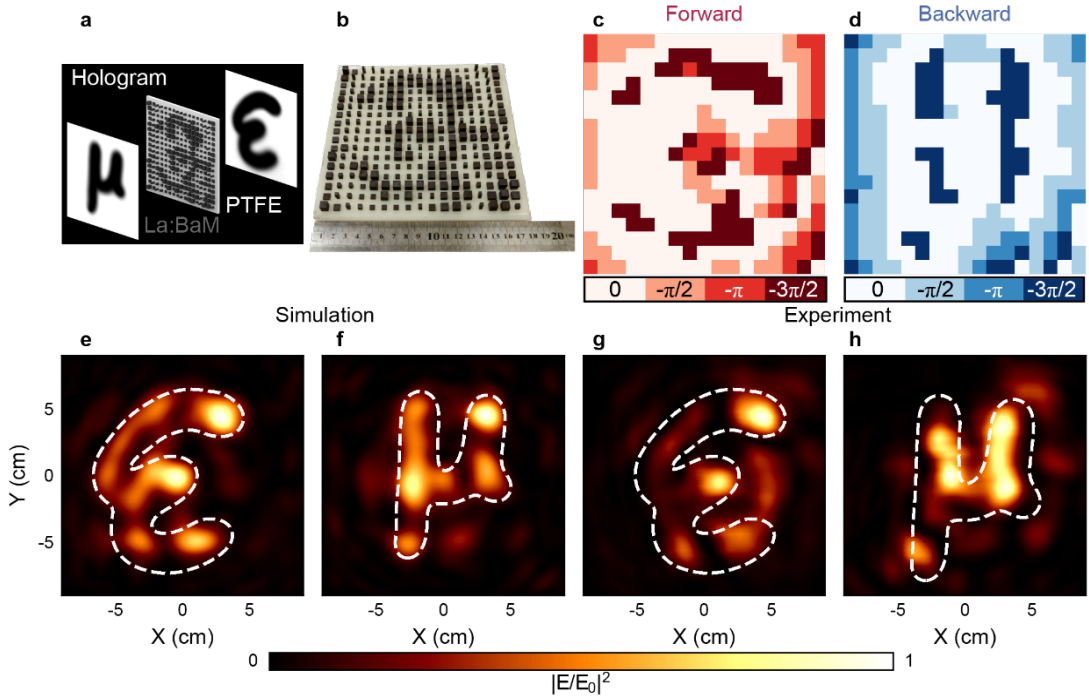

**Fig. 5| Nonreciprocal phase-only holograms. a**, Schematic of the nonreciprocal phase-only hologram sample. **b**, Image of the fabricated nonreciprocal phase-only hologram sample. Phase profiles of the metasurface for **c**, forward incidence and **d**, backward incidence. Simulated electric field intensity profiles of the metasurface for **e**, forward and **f**, backward incidence. Measured holographic images **g**, for forward and **h**, backward incidence.

## Conclusion

We proposed and fabricated a magnet-free nonreciprocal metasurface platform for on-demand bidirectional phase modulation. Owing to the strong magneto-optical effect and strong magnetocrystalline anisotropy of La:BaM, we achieved nonreciprocal and arbitrary intensity and phase profiles for both forward and backward propagation with subwavelength-scale Mie resonators. For the intensity-modulated NRM, we demonstrated one-way transmission in a wide angular range of 0° to ±64° at 15.7 GHz frequency. For phase-gradient NRMs, we demonstrated three functional devices: a nonreciprocal deflector, a nonreciprocal metalens, and nonreciprocal holography at the $K_u$ band. This technology can be potentially extended to cover the MHz to the optical frequency range by choosing appropriate self-biased magnetic materials. Magnet-free, low-profile, passive, linear, and broadband NRMs are poised to find broad applications in devices such as free-space isolators, nonreciprocal antennas, radomes, and full-duplex wireless communication links.

The platform demonstrated in this study overcomes several barriers that hamper the practical applications of NRMs. Self-biased magnetic meta-atoms eliminate the need for an external field, thereby resolving a major roadblock in the deployment of magnetic metasurfaces. Our approach also circumvents the challenges encountered in other nonreciprocal metasurfaces, such as the sensitivity to incident wave intensity, large active power consumption, high harmonic generation, low signal-to-noise ratio, poor power handling capability, and bandwidth limitations. Our metasurface designs are also insensitive to incidence angles, potentially enabling their application as omnidirectional antennas, for example, for wireless communications or integration on curved surfaces such as antenna radomes. The transmission efficiency of the metasurfaces reaches 30%–77% in the $K_u$ band, ranking among the best values in nonreciprocal metasurfaces reported so far[14, 35]. This is particularly remarkable considering that no gain elements were used. Even though the experiments were carried out only in the $K_u$ band, NRMs can be constructed in the V band (40 GHz to 75 GHz) or W band (75 GHz to 110 GHz) using the same La:BaM material by observing the nontrivial off-diagonal component at these frequencies in Fig. S2c. Furthermore, the operating frequency of such metasurfaces can be extended to cover the MHz to optical frequencies by choosing different self-biased magnetic materials. For example, in the MHz frequency range, cobalt or nickel spinel ferrites are ideal materials[36-39]; in the GHz to THz range, hexaferrites with tunable magnetic anisotropies are good candidates[40]; at optical frequency, self-biased garnet thin films using magnetoelastic effects are promising materials[41, 42]. Therefore, a variety of bias-free magnetic nonreciprocal metasurfaces are envisioned for future studies.

There is also considerable room for further improvement in device performance. First, reflection at the air-metasurface interface and optical absorption in the meta-atoms, on average, account for 15% and 40% efficiency losses, respectively, in the devices. Impedance-matching structures can be incorporated into meta-atoms to reduce the reflection loss. With optimized doping or growth protocols, the imaginary parts of $\varepsilon$ and $\mu$ in hexaferrite materials can be diminished[43]. Moreover, the large off-diagonal $\mu$ element implies a "non-perturbative" design scheme in which forward and backward waves yield different modal overlaps with meta-atoms, thereby suppressing the insertion loss to a level well beyond the material's figure-of-merit limit[44]. This effect is illustrated in Figs. S3i and S3j, where the scattering cross-section is different for forward and backward propagation in the same meta-atom, leading to a low insertion loss of only 1.1 dB in the one-way transmission metasurface. Second, polarization-independent nonreciprocal metasurfaces can be developed by leveraging different magneto-optical effects, such as transverse magneto-optical Kerr effects (TMOKE) for linearly polarized waves or the Faraday effect for circularly polarized waves. Third, the bandwidth of metasurfaces can be further extended. This is because the gyromagnetic property of the magnetic material is broadband, as shown in Fig. S2c. Dispersion engineering on both the forward and backward propagating modes can be performed on the meta-atoms, analogous to the design of an achromatic metalens, except for the bidirectional nature. These considerations will aid development of nonreciprocal metasurfaces for high-efficiency, broadband, and on-demand nonreciprocal control of electromagnetic radiation in the future.

## Methods

**Metasurface fabrication.** Metasurface fabrication involved the fabrication of La:BaM elements and their attachment onto a Teflon substrate. La:BaM bulk crystals were cut using an ultrahigh-precision computer numerical control (CNC) machine to fabricate square nanopillars with a dimensional accuracy better than 0.05 mm. Subsequently, the nanopillars were placed in an electromagnet with a maximum magnetic field of 2.3 T. The easy magnetization axis was saturated by gradually applying the voltage and current of the electromagnet, and the external magnetic field was gradually removed to maintain the remanent magnetization of the ferrite. The Teflon substrates were 18 cm × 18 cm sheets with a thickness of 0.5 cm. A double-sided adhesive was applied and the acrylic plate was covered with an array of pockets to facilitate meta-atom fixture. The magnetized meta-atoms were placed in predefined pockets and laminated with double-sided adhesive. After all elements were arranged in the substrate, the acrylic plate and unused double-sided adhesive were removed.

**Metasurface characterization.** All experimental results were measured using a vector network analyzer (VNA, Agilent E8363C) to obtain the $K_u$-band complex scattering parameters $S_{21}$ and $S_{12}$ in a microwave darkroom. To obtain the transmittance spectra of the metasurfaces, two identical circularly polarized antennas were placed on each side of the sample and connected to the two ports of the VNA via cables. Each antenna could function either as a source or detector to obtain the S-parameters. A Teflon lens with a focal length of 30 cm was placed between the metasurface and antennas on both sides to better focus energy on the sample. Calibration was first performed by eliminating the path loss without the sample using a two-port-through calibration. The sample was then placed to obtain the complex S-parameters. For circularly polarized antennas, only microwaves with the same handedness as the antennas could be detected.

Because the reflected microwave possesses opposite handedness to the incident wave, the $S_{11}$ and $S_{22}$ parameters vanish. Therefore, only two-port-through calibration was used instead of through-reflect-line calibration. To eliminate the multipath effect, the time-domain gate was activated in the VNA to smoothen the spectra.

For the transmittance spectra measurement under oblique incidence, the sample holder was replaced by a scaled rotatable platform (scale accuracy of 1°), and the scattering matrix under oblique incidence was obtained by rotating the sample.

For the beam deflector measurement, the sample and transceiver antenna were fixed on the same rotating platform with the rotation axis located at the center of the sample holder. The rotation angle was measured using an electronic angle ruler with 0.05° accuracy. The beam deflector sample was placed toward or opposite the transceiver antenna to obtain the intensity of the transmitted microwave. To fully characterize the radiation pattern, measurements were performed at angles ranging from 0° to 180°.

For the metalens and hologram measurements, only one Teflon lens was used in front of the source antenna to focus the incident microwave. The transceiver antenna (with a field resolution of 5 mm) was fixed on a linear translation stage driven by a stepper motor in the horizontal and vertical directions. The maximum moving range was 30 cm in both the horizontal and vertical directions. In our measurements, the moving speed of the receiving antenna was 5 mm/s. The custom-built motor scanning program communicated with the VNA to execute point-by-point measurements, and the energy/phase distribution on a specific plane was obtained.

**Metasurface simulation.** Simulations were performed using the commercial software COMSOL Multiphysics based on the finite element method. Periodic boundary conditions were employed in both transverse directions around each unit cell, whereas the polarization and incident angle of the microwave were defined in the incident port along the *z*- (longitudinal) axis. A perfectly matched layer was imposed on the incident port to absorb scattered waves. The phase and transmission spectra were inferred from the S-parameters in COMSOL. Home-made Python codes based on the Kirchhoff diffraction integral and Gerchberg–Saxton algorithm were used to calculate the nonreciprocal focusing characteristics and holographic patterns, respectively.

## Data availability

Source data are provided with this paper. All other data from this study are available from the corresponding authors upon reasonable request.

## Acknowledgements

The authors are grateful for the support from the Ministry of Science and Technology of the People's Republic of China (MOST) (Grant Nos. 2018YFE0109200, 2021YFB2801600), National Natural Science Foundation of China (NSFC) (Grant Nos. 51972044, 52021001, 52102357), Sichuan Provincial Science and Technology Department (Grant Nos. 2019YFH0154, 2021YFSY0016), Fundamental Research Funds for the Central Universities (Grant No. ZYGX2020J005), Open Foundation of Key Laboratory of Laser Device Technology, China North Industries Group Corporation Limited (KLLDT202102).

## Author contributions

L.B. and J.Q. conceived the idea. W.Y. (Weihao Yang) performed the device design, metasurface fabrication, and characterization. J.L., W.Y. (Wei Yan) and Y.Y. conducted the device characterization. C.L., E.L., L.D., and J.H. analyzed the data and wrote the manuscript. J.Q., L.D., Q.D., J.H., and L.B. supervised the study. All authors contributed to the technical discussions and writing of the manuscript.

## Competing financial interests

The authors declare no competing financial interests.